\documentclass[acmsmall,nonacm]{acmart}
\usepackage{multirow}
\usepackage{listings}
\usepackage{subcaption}
\usepackage[T1]{fontenc}
\usepackage{enumitem}
\usepackage{bbding}

\def\BibTeX{{\rm B\kern-.05em{\sc i\kern-.025em b}\kern-.08emT\kern-.1667em\lower.7ex\hbox{E}\kern-.125emX}}

\begin{document}
\title[RadGrad: Results of a case study]{Improving engagement, diversity, and retention in computer science with RadGrad: Results of a case study}

\author{Philip M. Johnson}
\email{johnson@hawaii.edu}
\author{Carleton Moore}
\email{cmoore@hawaii.edu}
\affiliation{%
  \institution{Department of Information and Computer Sciences, University of Hawaii at Manoa}
  \city{Honolulu}
  \state{HI}
  \country{USA}
  \postcode{96822}
}

\author{Peter Leong}
\email{peterleo@hawaii.edu}
\author{Seungoh Paek}
\email{spaek@hawaii.edu}
\affiliation{%
  \institution{Department of Learning Design and Technology, University of Hawaii at Manoa}
  \city{Honolulu}
  \state{HI}
  \postcode{96822}
  \country{USA}
}
\orcid{0000-0002-9508-9225}

\renewcommand{\shortauthors}{Johnson et al.}

\newcommand{\AbstractCategory}[1]{%
  \par\addvspace{.5\baselineskip}
  \noindent\textbf{#1}\quad\ignorespaces
}

\begin{abstract}

\AbstractCategory{Objectives}
RadGrad is a curriculum initiative implemented via an application that combines features of social networks, degree planners, individual learning plans, and serious games. RadGrad redefines traditional meanings of ``progress'' and ``success'' in the undergraduate computer science degree program in an attempt to improve engagement, retention, and diversity. In this paper, we describe the RadGrad Project and report on an evaluation study designed to assess the impact of RadGrad on student engagement, diversity, and retention. We also present opportunities and challenges that result from the use of the system.

\AbstractCategory{Participants}
Our evaluation study involved four major stakeholder groups for undergraduate computer science education at our institution: students in the early stages of the degree program, recent graduates, faculty, and advisors.

\AbstractCategory{Study Methods}
We used a qualitative experimental research design, including data obtained through questionnaire responses from our major stakeholder groups.

\AbstractCategory{Findings}
We found strong evidence to support the hypothesis that RadGrad can have a positive impact on student engagement in Computer Science for students who are relatively new to the discipline. We found some evidence to support the hypothesis that RadGrad can have a positive impact on student retention. We were unable to evaluate the impact of RadGrad on diversity.

\AbstractCategory{Conclusions}
The data gathered in this study indicates that approaches like RadGrad provide a promising means to improve engagement, retention, and (potentially) diversity in Computer Science.  The data suggests that development and evaluation of RadGrad at other institutions and/or in other STEM disciplines would provide useful insight into the generality of the approach. It also indicates that faculty engagement with the technology might increase its effectiveness, but that obtaining faculty buy-in is challenging.

\end{abstract}



\maketitle

\section{Introduction}

\subsection{Research Problem}

Traditional curriculum initiatives for undergraduate computer science generally fall into one of three categories. The first involves the injection of a specific technology across the curriculum, such as initiatives for cloud computing \cite{deb_module-based_2019}. The second involves the injection of a specific domain across the curriculum, such as initiatives for cybersecurity \cite{tang_shaping_2019} or distributed computing \cite{abebe_watdfs:_2019}. The third involves the integration of practices intended to address demographic problems including lack of access or diversity, such as initiatives at the University of Illinois \cite{metcalf_diversity_2018} and University of Oklahoma \cite{collain_you_2019}.

One problem with these types of curriculum initiatives is that they all require changing, well, the curriculum.  This is a problem because changing the curriculum is hard. That is, curriculum changes can present challenges for both academics and their communities for several reasons, such as resisting strong top-down control and seeking meaningful justifications for changes \cite{annala_understanding_2021}. In addition any significant curriculum change will require the approval of committees at multiple levels of the educational institution, it will require buy-in across the faculty, and it will often require significant resources to implement and sustain.

A benefit of changing the curriculum is that if it is accomplished, then {\em student} buy-in is normally not required. Because the change is to the curriculum, if a student wants to graduate, then they must follow the changed curriculum.

At our University, we have been experimenting with an initiative called "RadGrad", which falls outside these typical categories of curriculum initiatives, and as a result inverts the typical problems and benefits associated with traditional curriculum initiatives.

RadGrad originally rose from two insights: first, a recognition that our department did not have the resources to enable its undergraduate curriculum to keep up with the rate of change in computer science technology and the ever widening domain of application areas, nor could we keep up with student demand for access to curriculum content.  In some cases, the local high tech community or online educational services could fill in the gaps, but our department provided no support for student awareness of these extracurricular resources, or guidance as to which ones were worth spending time on, or how such activities might integrate with existing curricular offerings.  Finally, since extracurricular activities, are, by definition, "extra", there is a structural incentive for students to limit these activities and instead focus on curricular work which directly affects their GPA.

Second, we recognized that the increase in demand for our undergraduate computer science degree programs was paradoxically leading to a decrease in the diversity of our student body. For example, the representation of women in computer science is low in the United States and has been declining over the past few decades \cite{moudgalya_computer_2019}. According to the U.S Department of Education \cite{us_department_of_education_digest_2019}, the percentage share of BS degrees earned by women is approximately 20\% in 2018-2019, which is down from over 35\% in early 1980s. These statistics are also similar to our program. That is, increasingly few women and underrepresented groups were making it all the way through our program to graduation. Although approximately 40\% of our first year students each year are women, only 23\% of our graduating seniors in AY 2021 were women.

In an attempt to address these issues, we created RadGrad, which is a web-based application that combines features of social networks, degree planners, and serious games.  Undergraduate students in our introductory courses are provided an account on RadGrad, where they can declare their interests and career goals. The system then recommends courses and extracurricular opportunities matching their preferences. A degree planner tool enables students to lay out their "degree experience" as a combination of curricular activities (courses) and extracurricular activities (internships, meetups, online courses, hackathons, etc) for each upcoming semester until their planned graduation date. Finally, RadGrad enables students to learn about the broader ways in which computer science impacts upon local, national, and global society, as well as helping them to connect with and create communities of practice inside and outside of the department. As will be discussed below, related research suggests that making these connections can improve student engagement, diversity, and retention.

To combat the view of extracurricular activities being perceived as "extra", RadGrad does not use GPA to represent student success or progress. Instead, RadGrad provides a three component metric called "myICE", an acronym for Innovation, Competency, and Experience. Students are awarded Competency points for successfully completing a course, and a varying amount of Innovation and/or Experience points for successfully completing an extracurricular activity defined within the system. For example, participating in a hackathon will earn some number of Innovation points, since the goal of hackathons is to create something new.  Finishing a summer internship in a high tech company will earn Experience points since that activity provides students with a sense for the demands of a high tech workplace environment. (Depending upon the nature of the internship, it could also earn Innovation points.) To "win" at RadGrad, students must (among other things) earn at least 100 Innovation, Competency, and Experience points by completing some combination of curricular and extracurricular activities by the time they graduate.

Unlike traditional curriculum initiatives, RadGrad exists apart from the official curriculum requirements for our degree programs. As a result, we have been free to deploy and experiment with the approach without the multi-year approval process normally required for curriculum changes.  On the down-side, participation in RadGrad is voluntary, and so students must opt in to the system and its philosophy to gain its benefits.

\subsection{Research Goals}

RadGrad began in the Fall of 2015 as a student-driven project in several upper-division software development courses.  A common perspective from students working on the project at that time was, "I wish I'd had RadGrad when I was starting out in Computer Science".  However, most of these students were high achieving, soon-to-graduate individuals who were demonstrating an ability to successfully finish a computer science degree program without the aid of RadGrad.  While we found their feedback to be encouraging, we wanted to know whether RadGrad would be interesting and/or useful to students who were more "at risk" in our program. Ideally, we wanted to know if the insights that RadGrad provides regarding the spectrum of computer science careers as well as extracurricular opportunities in the discipline both during and after the degree program would lead to a more diverse and engaged student body, with less attrition over the course of the degree program.

Unfortunately, there have been a multitude of impacts on our computer science degree program over the years of the RadGrad Project, ranging from curricular revisions to changes in advising personnel and process to the biggest impact of them all: COVID-19.  Our research approach cannot unambiguously separate the impact of RadGrad on engagement, retention, and diversity from all of these other factors.  While we will report on institutional data regarding diversity and retention, our approach focuses on qualitative perspectives from four stakeholder groups (students in the early stages of the degree program, recent graduates from the degree program, faculty, and advisors) on their view of RadGrad and the use of this data to assess the potential of RadGrad to impact engagement, diversity, and retention, along with insights into how to improve the project in the future.

More specifically, during Academic Year 2021-2022, we conducted a qualitative case study that was designed to provide evidence regarding the following research questions:

\begin{enumerate}
\item Do early stage students, when introduced to RadGrad, find it of value to their educational experience?
\item Do recent graduates, having just completed the degree program, view RadGrad as having improved their educational experience?
\item What are the views of Faculty regarding RadGrad?
\item What are the views of Advising staff regarding RadGrad?
\item What do these findings imply regarding engagement, diversity, and retention in computer science undergraduate degree programs?
\end{enumerate}

We report on this study starting in Section \ref{sec:method}. First, however, we step back to better motivate the design of RadGrad, as well as describe the system in more detail.

\section{Related Work}
\label{sec:related-work}

There is a national need for undergraduate computer science degree programs to improve both retention (the percentage of students entering CS programs who finish the degree) and diversity (the percentage of graduates who are female and/or from an underrepresented minority group).  We need to improve retention because the projected demand for skills in computer science far exceeds current production \cite{camp_generation_2017}.  We need to improve diversity because a more diverse STEM population improves tech innovation at large. For example, mixed-sex teams filed 40\% more information and technology patents than all-male teams \cite{ashcraft_who_2012}, and management diversity leads to a \$42M increase in S\&P value of firms \cite{dezso_girl_2007}.

While the need is clear, solutions are complicated. Gender diversity in computer science has actually fallen in the last 20 years \cite{hong_women_2014}, with no well accepted explanation for its cause. Some diversity-related issues start in middle and high school: black students are less likely than white students to have computer science courses in middle and high school, and female students are less likely than male students to be told they would be good at computer science \cite{inc_diversity_2016}.   There is some research that provides evidence for a way forward: a study by Google \cite{hong_women_2014} concludes that four factors primarily influence young womens' decision to pursue CS: (1) social encouragement (positive reinforcement of CS pursuits from family and peers); (2) self perception (an interest in problem solving and a belief that those skills can be translated to a successful career); (3) academic exposure (availability of curricular and extracurricular CS activities); and (4) career perception (view of CS as a career with diverse applications and a broad potential for positive societal impact). Stout and Camp \cite{stout_now_2014} make similar points around social relevance, a sense of belonging, and cultural bias. RadGrad implements capabilities designed to help address isssues around self perception, academic exposure, career perception, and social relevance among its student users.

For those high school students who graduate and enter an undergraduate degree program in computer science, retention becomes a significant issue.   More than half of the students who start out in science or engineering switch to other majors or do not finish college at all \cite{kober_reaching_2015}. Initiatives to improve retention, such as the Threads undergraduate curriculum at Georgia Tech, emphasize giving students more control over their degree plan, a better understanding of how their studies relate to their career interests, and an increased emphasis on the importance of extracurricular activities \cite{barrett_expanding_2017}. RadGrad provides a technology platform, information system, and incentive structure with these emphases.

Communities of Practice (CoP) is a theory of learning first proposed in 1991 \cite{lave_situated_1991}, more fully developed in 1998 \cite{wenger_communities_1998}, and extended to "landscapes of practice" in 2004 \cite{wenger_learning_2004}. A loose definition of CoP is "groups of people who share a concern or a passion for something they do and learn how to do it better as they interact regularly." More specifically, three characteristics distinguish a CoP from other kinds of communities: (1) There is at least one domain of interest shared by all members; (2) members engage in joint activities and discussions, help each other, and share information;  and (3) members are practitioners in the domain, not just people with shared interests, and thus develop a shared repertoire of resources.

Communities of Practice show great promise for improving undergraduate retention \cite{barker_results_2014} and diversity \cite{gardner_authentic_2015}, because participating students will find a new source for social encouragement, self-perception, academic exposure, and career perception. In undergraduate degree programs, CoPs are primarily found within disciplinary-specific extracurricular activities: clubs, meetups, hackathons, and so forth. RadGrad makes these CoPs visible to students, and provides incentives to students for participating in them.

\section{RadGrad}
\label{sec:radgrad}

The RadGrad project began in the Fall of 2015. For the first few years, development was student-led as part of various software engineering classes. Development activities during this time focused on the underlying data model, user interface mockups, and possible game mechanics. In 2017, a pilot implementation was deployed into a computer science department and some we began receiving user feedback. In 2018, we began work on RadGrad Version 2, an "industrial strength" version of the system that followed high quality software engineering practices and would enable deployment to different institutions and disciplines. From a technical perspective, RadGrad is now a functional and reliable web-based application, implemented in approximately 30,000 lines of Javascript and 7,000 lines of HTML. It has extensive design and development documentation, unit and integration tests, and implements continuous integration. Instances can be deployed locally or using a cloud-based hosting service such as Digital Ocean. Over 1,000 students have used RadGrad to date.

\subsection{Theory of Change}

\begin{figure}[ht]
\centering
\includegraphics[width=\linewidth]{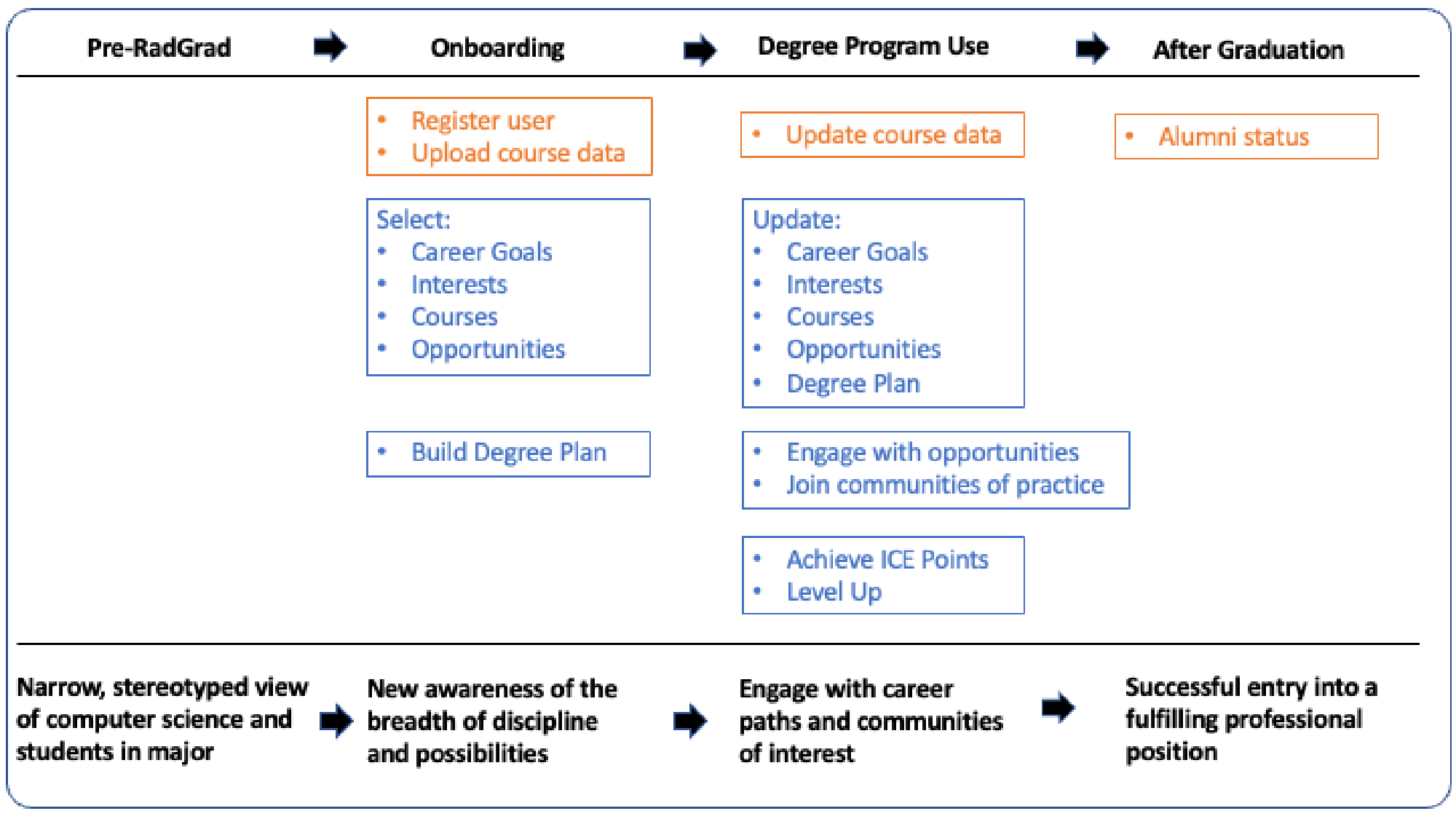}
\caption{\em RadGrad Theory of Change/Workflow. Orange boxes indicate actions taken by administrators, blue boxes indicate actions taken by students.}
\label{fig:theory-of-change}
\end{figure}

These experiences have helped us understand the "workflow" of RadGrad, and how it implements a theory of change for students, as illustrated by Figure \ref{fig:theory-of-change}.

We find that many students in the {\em Pre-RadGrad} phase have a narrow, stereotyped view of computer science as a discipline (i.e. "it's only about video games") and the students who pursue the major (i.e. "introverted, anti-social white or asian male gamers").

The {\em Onboarding} phase represents the first exposure of students to RadGrad, and hopefully occurs by the time of their first or second semesters in the degree program. Each RadGrad instance is invitation-only, so an administrator must run scripts to create accounts and upload data on completed degree program coursework. Once registered, students login to the system using their institutional authentication mechanism. We provide an online tutorial that students can follow to familiarize themselves with the system, which teaches them how to select an initial set of Career Goals, Interests, Courses, and Opportunities that they find of interest. RadGrad provides a recommendation system so that once students select Career Goals and Interests, the system can bring related Courses and Opportunities to their attention.  The anticipated outcome of the Onboarding process is an increased awareness of the range of disciplinary pursuits, along with concrete activities to pursue to actively explore them.

The {\em Degree Program Use} phase represents the remainder of the student's degree program, in which they can refer back to RadGrad for new Career Goals and Interests and how they relate to new Courses and Opportunities. RadGrad also includes a Review system that enables students to read about the experiences of other students with Courses and Opportunities. Unlike conventional online systems (like "Rate My Professor"), RadGrad Reviews are not anonymous and are subject to moderation; these two features encourage high quality, thoughtful reviews.  The most important goal during Degree Program Use is for students to actually take part in Opportunities, which will connect them with communities of practice (both on or off-campus) in a particular disciplinary area. RadGrad provides two game mechanics (myICE Points and Levels) for students who are motivated by those kinds of things.

The {\em After Graduation} phase represents the state of affairs shortly after matriculation.  Our theory of change predicts that students who have taken part in RadGrad will have been more engaged with a wider variety of disciplinary activities occurring both on and off campus, and that those experiences will help them successfully enter into a fulfilling professional position. Once a student graduates, a RadGrad administrator changes their role to "Alumni", which currently converts their account to read-only status.

\subsection{From theory of change to user interface}

Designing a desirable theory of change is one thing, designing and implementing a usable UI to support it is quite another. We redesigned the RadGrad user interface several times over the past five years based on feedback from students, faculty, and advisors.  RadGrad currently provides specialized user interfaces for each of five roles: Student, Faculty, Advisor, Admin, and Alumni. Providing the appropriate UIs for all of these roles requires approximately 130 different web page designs. In this section, we present just a few pages to provide the flavor of RadGrad's user interface.

\subsubsection{Home Page/Checklist}

We learned from our early prototypes that the "conceptual model" implemented by RadGrad is not intuitively obvious. RadGrad users had problems understanding both what to do and why to do it.  We eventually discovered a way to address this "disorientation" by making the Home Page a "smart checklist" with a prioritized list of things to do.  Figure \ref{fig:radgrad-student-home-page} illustrates a portion of the home page for a hypothetical student:

\begin{figure}[ht]
\centering
\includegraphics[width=\linewidth]{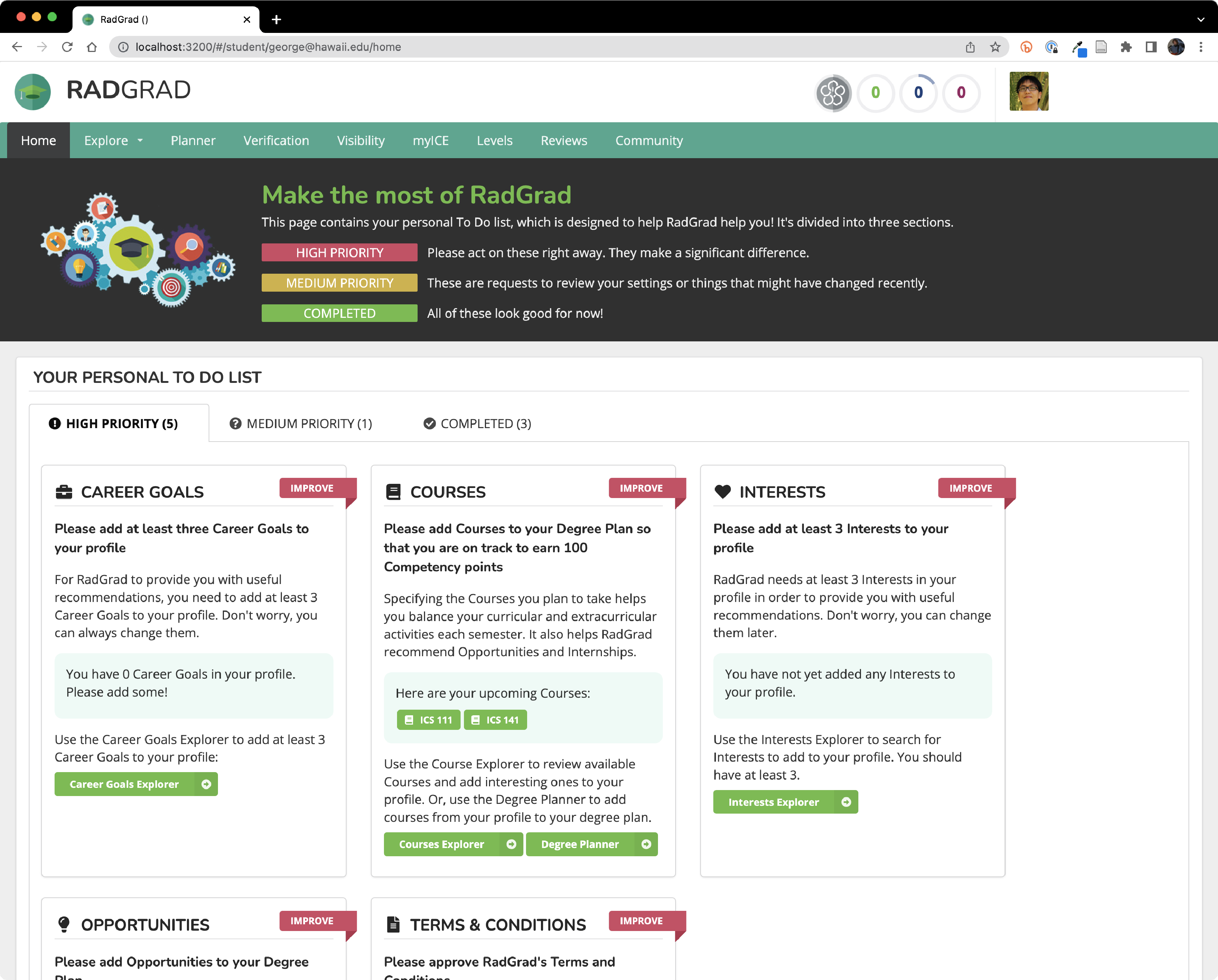}
\caption{\em RadGrad Home Page (Student Role)}
\label{fig:radgrad-student-home-page}
\end{figure}

Each time this page is retrieved, all of the checklist items associated with the role of the user run code to check various state values and decide whether the checklist item should be displayed as "High Priority", "Medium Priority", or "Completed", as well as what should be included in the content of the checklist item.

This checklist design has been well received by users, because it provides direct, personalized feedback on both what to do next (if anything), as well as why to do it. Checklists are not only useful when a user is first learning to use RadGrad, but is also helpful as a means to inform users of new content in the system or other changes.

Finally, checklists are helpful not just to students!  We find that faculty, advisors, and even administrators benefit from a smart, personalized checklist home page.

\subsubsection{Content Explorer Pages}

The current version of RadGrad for computer science at our institution defines 20 Career Goals, 78 Interests, 57 Opportunities, and 106 Courses. This is a significant amount of inter-dependent content.  RadGrad provides "Explorers" for each of these four content types that enable users to scroll through brief summaries of each item.

\begin{figure}[ht]
\centering
\includegraphics[width=\linewidth]{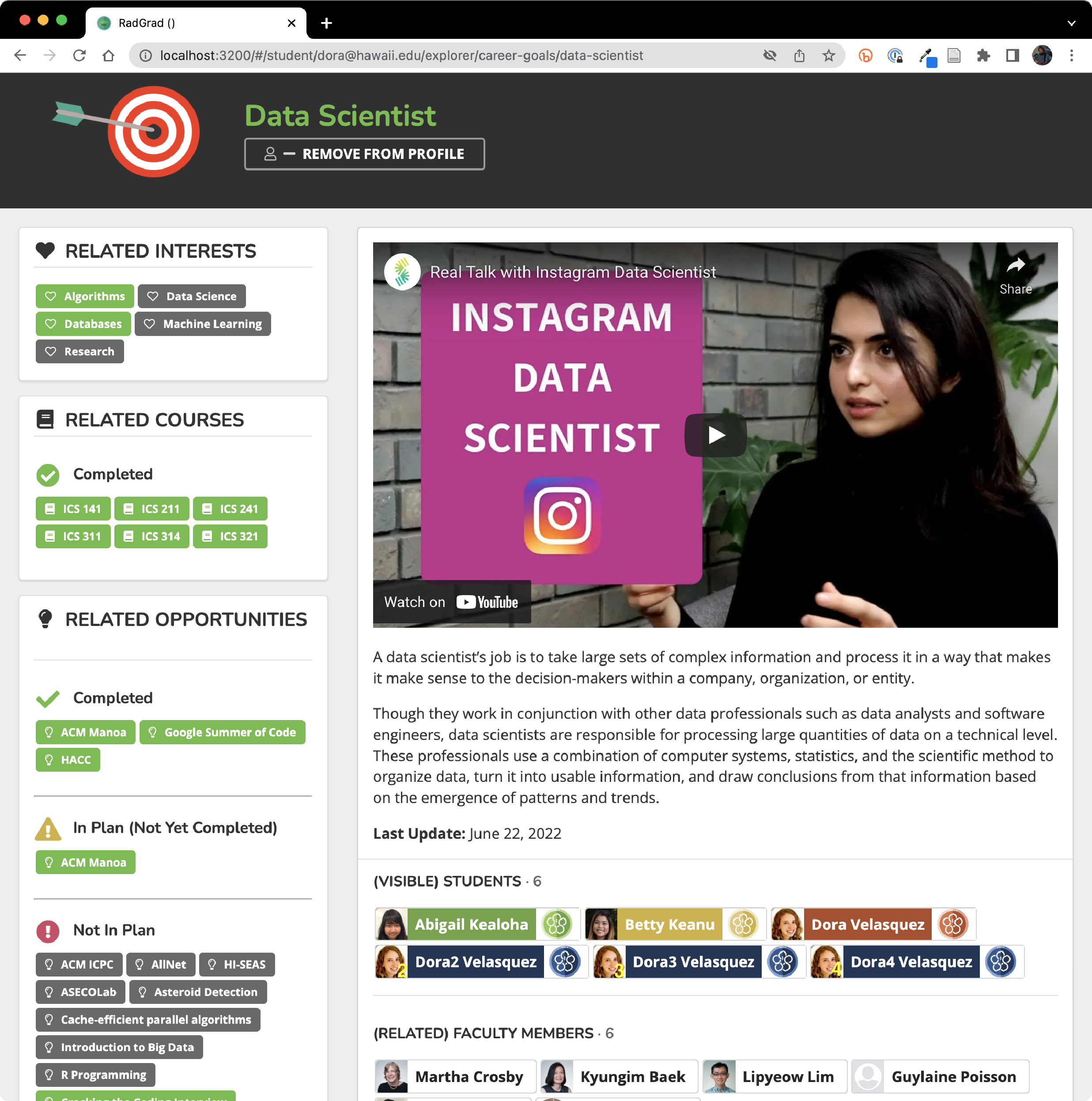}
\caption{\em RadGrad Data Scientist Details Page. (Student names are pseudonyms.))}
\label{fig:radgrad-data-scientist-detail}
\end{figure}

It is straightforward to search the Internet for "Computer Science Careers" (for example) and find a plethora of pages. RadGrad provides unique value to students by providing relationships between content items and users, as illustrated in Figure \ref{fig:radgrad-data-scientist-detail}.

This page shows that for this (hypothetical) student user, (a) they have added Data Scientist to their profile as a Career Goal; (b) Of the five Interests associated with the Data Scientist Career Goal, the student has two of them in their profile (Algorithms and Databases); (c) The student has completed six Courses related to Data Science already; (d) The student has already completed three Opportunities related to Data Scientist, has one upcoming, and there are quite a few more related Opportunities that could be added to their plan; (e) Six students in this community have Data Scientist in their profile (and have opted in to make this information visible to other registered users), and (f) six faculty have listed Data Scientist as a Career Goal for which they could provide advice and guidance.  All of this information is in addition to a brief profile of the Career Goal along with a "Day in the Life" video.

It is important to note that RadGrad content is designed to be "faculty-curated".  It is the responsibility of one or more faculty in the discipline at the institution to determine what constitutes relevant Career Goals, Interests, and Opportunities for their students, and how these items are documented in the system. There is no central authority decreeing content for all instances of RadGrad. This design enables a RadGrad instance to be appropriately tailored to the educational context, but implies ongoing faculty participation in content development.

\subsubsection{The Degree Plan Page}

\begin{figure}[ht]
\centering
\includegraphics[width=\linewidth]{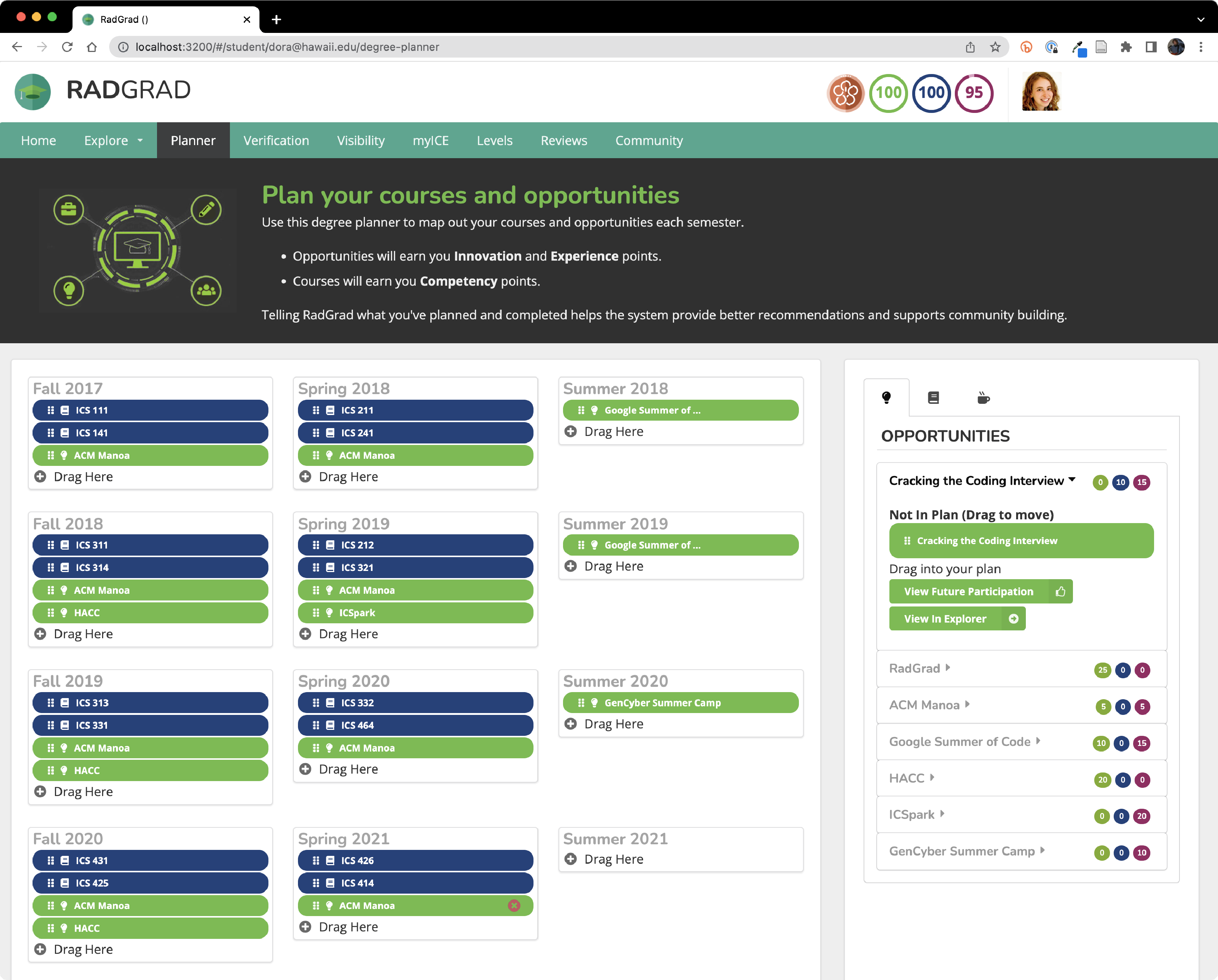}
\caption{\em RadGrad Degree Plan Page (Student Role)}
\label{fig:radgrad-student-degree-plan}
\end{figure}

As part of the onboarding process, students peruse RadGrad content and add relevant Career Goals, Interests, Opportunities, and Courses to their "profile".  Once a student has identified {\em what} they are interested in, the next step is to determine {\em when} they want to participate. Answering the latter question is the function of the Degree Planner, as illustrated in  Figure \ref{fig:radgrad-student-degree-plan}.

The left side of the Degree Planner contains a grid representing a set of academic years, each containing three semesters (Fall, Spring, Summer)\footnote{RadGrad can be configured to support institutions on the quarter system.}.  For each semester, the student can specify the Courses and Opportunities in which they plan to participate.

The right side provides access to the Courses and Opportunities in the student's profile. By expanding an item (such as the "Cracking the Coding Interview" Opportunity), a tile is revealed that can be dragged to any current or future semester.

\subsubsection{Levels and myICE: What's good enough?}

So far, we've presented a design that enables students to peruse faculty-curated content regarding the discipline, find curricular (Courses) and extracurricular (Opportunities) related to their Interests and Career Goals, and plan out when they want to participate.

But there's two significant issues:
\begin{enumerate}
\item Sure, the more courses and experiences one has, the better, but can we do better? Can we help students to understand what is "enough"?
\item Does RadGrad just dump more "work" on students? Can we inject a little fun into the process?
\end{enumerate}

\begin{figure}[t]
\centering
\includegraphics[width=2in]{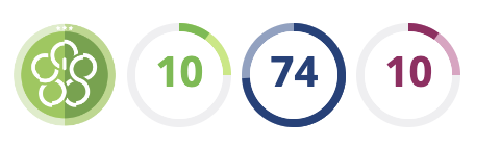}
\caption{\em UI showing the student Level and their Innovation, Competency, and Experience myICE points. This student is at Level 3 (i.e. Green). They have planned 25 Innovation points, and earned 10. They have planned (at least) 100 Competency points, and earned 74. They have planned 25 Experience points, and earned 10.  }
\label{fig:ice}
\end{figure}

We designed two game mechanics, myICE points and Levels, to address these issues.  myICE points provide a way for students to "complete" a degree plan, in the sense that if they complete all of the Courses and Opportunities in a complete degree plan, they will be both "well prepared" and "well rounded".  Levels provide a graphical illustration for a student's progress through the degree program. Figure \ref{fig:ice} shows the portion of the navigation bar containing both the student's Level (in this case, Level 3, represented by a green RadGrad icon) and their current myICE Points (three circles containing the numbers 10, 74, and 10).

Let's start with myICE points.  myICE points defines a metric for the undergraduate degree experience that is independent from grades (and thus the GPA) and also elevates extracurricular activites (i.e. Opportunities) to first class status along with Courses. myICE defines three submetrics: "Innovation", "Competency", and "Experience", and students earn points in one or more of these categories each time they complete a Course or Opportunity. In the computer science RadGrad instance at our institution, students earn around 10 Competency points for completing a Course, and between 5-25 Innovation and/or Experience points for completing an Opportunity.

In addition to defining and describing the set of Opportunities in a RadGrad instance, it is also up to the faculty to decide which ICE points, and how many, a student should earn for each Opportunity's completion. For example, in the computer science RadGrad instance at our institution, participating in the student chapter of a professional society earns 5 Innovation and 5 Experience points per semester.  Participating in a research group earns 25 Innovation points per semester.  A summer internship at a high tech company earns 25 Experience points.

The goal is to assign points such that students who have earned 100 Innovation, 100 Competency, and 100 Experience points would be considered both "well rounded" and "well prepared" upon graduation.  While this point system is very heuristic in nature, it is still more descriptive, and prescriptive, than the only institutional metric currently available to students: GPA.

To distinguish "planned" from "earned" ICE points, RadGrad includes a verification process for both Courses and Opportunities. For Courses, RadGrad imports data from the University's course management system (STAR) indicating the coursework the student has taken and the grade they received.  To verify completion of an extracurricular activity, the student must manually "request verification" of the Opportunity.  This request sends a message to a RadGrad administrator or faculty member who has responsibility for checking that the student participated. The admin or faculty member can accept or decline the verification request, and if accepted, the system will award the points associated with the Opportunity to the student.

RadGrad represents myICE by three circles: a green circle for Innovation, a blue circle for Competency, and a red circle for Experience. For each of the three components of myICE, the circle is partially or fully colored in a light shade to represent the number of planned points, based upon what the student has put in their degree plan. If the student has earned the points, then a dark color is used to fill in part or all of the circle. The circles only represent 100 points for each of the three metrics; it is up to faculty to define points for Courses and Opportunities such that earning 100 points suffices to indicate a well prepared and well rounded student.

For example, in Figure \ref{fig:ice}, the Innovation circle is colored only a quarter of the way in a light green color, which indicates that this student has only planned to earn 25 Innovation points so far. One tenth of the circle is colored in a dark green color, indicating that 10 points have been earned. The blue circle is completely filled in, indicating that the student has planned to achieve 100 Competency points, and 74\% of the circle is dark blue, indicating 74 earned Competency points.  Finally, the Experience circle is only a quarter filled, indicating only 25 Experience points are currently in the degree plan, and 10\% of the circle is filled in a dark red color, indicating 10 earned Experience points.

Let's now look at Levels.  RadGrad provides six "Levels", each represented by a different colored RadGrad icon. As students earn myICE Points and carry out other RadGrad activities, such as writing reviews of courses and opportunities, they can progress through white, yellow, green, blue, brown, and black levels, similar to martial art belt colors. The rules for achieving Levels are designed such that a motivated student can achieve a new Level each semester, so a total of six semesters (typically three years) is required to get through all Levels.  The student in Figure \ref{fig:ice} is currently at Level 3 (green).

In summary, myICE points and Levels are designed to address the issues of "How much is enough?" and (to some extent) "Are we having fun yet"?

\subsection{From user interface (UI) to user experience (UX)}

"User interface" refers to what the user sees and interacts with on their screen. "User experience" includes the user interface but includes non-UI factors relevant to system adoption and use.  In RadGrad, we've found that the user experience beyond the interface is very important. Here are three significant aspects of the RadGrad user experience.

First, we wanted a way to make RadGrad visible "in real life".  To do this, we designed and printed laptop stickers corresponding to each of the six Level icons. When students reach a new level, they can go to an Advisor and get the laptop sticker corresponding to their new Level. We hoped that students would put RadGrad stickers on their laptop, and that this would provide physical advertisement of the system, and also potentially serve as a kind of status symbol for students who attain higher Levels in RadGrad.

Second, we experimented with two different approaches to introducing students to RadGrad. Initially, we obtained permission from instructors of introductory courses to attend a class session and step students through the onboarding process (i.e. selecting an initial set of Interests and Career Goals, then adding an initial set of Courses and Opportunities to their Degree Plan).  In-class onboarding had the advantage of engaging all the students in RadGrad simultaneously, and also provided us with an opportunity to hand out RadGrad laptop stickers to everyone based on their Level at the end of the class period.  But, we found that the 30 minutes spent introducing students led to a "rushed" experience, and did not allow students to explore the system and its content at their own pace.

Once COVID-19 hit, we switched to an online New Student Tutorial that provided a more comprehensive introduction than our in-class presentation, and allowed students to work their way through the system at their own pace.  To incentivize students to onboard, this activity was provided as an extra credit assignment in introductory courses.  The assignment also required students to write a short essay which was used as data for this evaluation.

Third, we asked advisors to integrate RadGrad into the student advising sessions as a way to support longer-range career and professional goal discussions.  As we found in our evaluation, student perceptions of faculty and advisor engagement with RadGrad were important in their perception of the utility of the system.

\section{Evaluation Method}
\label{sec:method}

\subsection{Research Design Overview}

In the best of all possible worlds, we would evaluate our research questions by a controlled, randomized experiment where half of our students would constitute the experimental "RadGrad" treatment, and the other half would function as the control group.  Our selection process would involve blocking in order to ensure that there were equal numbers of women and underrepresented groups in both groups, we would involve sufficient numbers of students to obtain statistically significant results, and we would follow them both during the degree program and following graduation.

This design was not feasible for several reasons: pilot study responses to RadGrad from students were so positive that we did not want to restrict access; we have low numbers of women in our program, making it difficult to obtain a large enough sample size to generate statistically significant differences in outcome data, it would require advising staff to know which group a student was in and alter their advising session appropriately, and tracking students post-graduation is difficult.

To both fit the constraints of our situation and still obtain useful data, we instead designed and implemented a qualitative case study. Our approach involved gathering a variety of data in a variety of formats from four stakeholder groups during Academic Year 2021-2022: (a) students in the early stages of their degree program, just after completing the onboarding experience ("Early Stage Students"); (b) students who had just completed (or were just about to complete) the degree program ("Recent Grads"); (c) faculty members ("Faculty"); and (d) student advisors ("Advisors").

Figure \ref{fig:eval-questions} summarizes how our "high-level" research questions (the impact of RadGrad on engagement, retention, and diversity) relates to the "low-level" questions for which we actually gathered data. A "?" in a cell indicates that the data regarding that low-level question might provide insight into the high-level question.

\begin{figure}[th]
\centering
\small
\begin{tabular}{ p{1in} p{1in} c c c }
\hline
\multicolumn{2}{l}{Does RadGrad have a positive impact on...} & Engagement? & Retention? & Diversity? \\
\hline\\[0.05cm]
Are the following features useful... & &  &  & \\
\multicolumn{2}{r}{Explorers?} & ? & ? & \\
\multicolumn{2}{r}{Degree Planner?} & ? & ? & \\
\multicolumn{2}{r}{Game Mechanics?} & ? &  & \\
\multicolumn{2}{r}{Reviews?} & ? & ? & \\[0.5cm]
RadGrad is... & &  &  & \\
\multicolumn{2}{r}{well designed?} & ? &  & \\
\multicolumn{2}{r}{useful for women and underrepresented groups?} &  & ? & ? \\
\multicolumn{2}{r}{helpful for creating communities of practice?} & ? & ? & ? \\
\multicolumn{2}{r}{helpful for broadening student perspectives on CS?} & ? & ? & ? \\
\multicolumn{2}{r}{useful for faculty?} & ? & ? & \\
\multicolumn{2}{r}{useful for advisors?} & ? & ? & \\
\hline
\end{tabular}
\caption{\em Research questions and the case study evaluation data that provide evidence regarding them.}
\normalsize
\label{fig:eval-questions}
\end{figure}

Appendices \ref{sec:onboarding-student-essay-prompt}, \ref{sec:graduating-student-questionnaire}, \ref{sec:advisor-questionnaire}, and \ref{sec:faculty-questionnaire} show the instruments used to collect data. As the Appendices and Figure \ref{fig:eval-questions} indicate, we did not ask participants questions that directly address our research questions, such as: "Does RadGrad make you feel more engaged with computer science?", "Do you feel more inclined to continue the computer science degree program because of RadGrad?", or "As a woman or member of an underrepresented group, does RadGrad make you more interested in computer science?".  This is because we do not believe direct questions like this would yield meaningful answers.

Instead, we asked a combination of direct questions about specific features of RadGrad (such as, "Did you find the course explorer to be useful") for graduating students, faculty, and advisors, along with an extremely open ended question for new students ("Write a 2-3 paragraph essay about RadGrad").  In the case of the open-ended essay, we then coded the essay to extract recurrent themes.

This combination of qualitative data results in a rich source of insight into the RadGrad experience across the four stakeholder groups. As will be shown in Figure \ref{fig:eval-questions-evidence}, it provides indirect evidence regarding all three of our top-level research questions, as well as useful and actionable information about the RadGrad UX.

\subsection{Study Participants}

\subsubsection{Researcher Description}

The research team consists of two Professors from the Department of Computer Science, and two Professors from the Department of Education.

\subsubsection{Participants}

Participants were drawn from four stakeholder groups:

\begin{enumerate}
\item {\em Early stage students.}  This stakeholder group consisted of students in the first, second, or third semester of the degree program during Academic Year 2021-2022.  Out of 475 students in this group, 187 participated, for a response rate of 39\%.

\item {\em Recent graduates.}  This stakeholder group consisted of students who completed the degree during in Fall 2021 or Spring 2022. Out of 107 students in this group, 23 participated, for a response rate of 21\%.

\item {\em Faculty.} This stakeholder group consisted of faculty in the degree program, excluding the two Professors who are researchers in this project.  Out of 21 faculty in this group, 10 participated, for a response rate of 47\%.

\item {\em Advisors.} This stakeholder group consisted of the two academic advisors in the degree program.  Both participated, for a response rate of 100\%.

\end{enumerate}

\subsubsection{Researcher–Participant Relationship}

There were two relationships between researchers and participants that could have impacted on the data. First, the researchers who are Computer Science Professors were also instructors for several of the courses being taken by the undergraduates in the early stages of their degree program. If students were aware that the instructors were also members of the RadGrad research team, that could potentially have influenced their responses.

Second, the researchers who are Computer Science Professors are also faculty in the Computer Science department. So, there are professional and personal relationships that could potentially influence the responses of faculty to their questionnaire.

\subsection{Participant Recruitment and Selection}
\subsubsection{Recruitment Process}

Participants for all stakeholder groups were contacted via email.  For undergraduates in the early stages of our degree program, they received a small amount of extra credit for completing the assignment.  This experiment was approved by our institution's institutional review board.

All participants who provided responses are included in the study.

\subsection{Data Collection}

\subsubsection{Data Collection/Identification Procedures}

We identified the {\em Early Stage Student} stakeholder group as all students in the first semester programming course, second semester programming course, and third semester software engineering course during Academic Year 2021-2022. We collected data from them in the form of an open ended essay written upon completion of the New Student Tutorial, which serves to onboard students to RadGrad. The essay prompt is provided in Appendix \ref{sec:onboarding-student-essay-prompt}.

During the New Student Tutorial, students are guided through the completion of several tasks, including: (a) Logging in to the system, (b) learning about the "smart checklist" on their home page; (c) using the Interest, Career Goals, Courses, and Opportunities explorers and adding items of interest to their Profile; (d) using the Degree Planner to lay out Courses and Opportunities across future semesters; (e) understanding myICE and Levels. As part of the tutorial, students are asked to "complete" their Degree Plan, which means finding enough Courses and Opportunities (corresponding to their Interests and Career Goals) and adding them to their Degree Plan so that, if they actually carry out the Courses and Opportunities, they will earn at least 100 Innovation, 100 Competency, and 100 Experience points by the time they graduate.

We identified the {\em Recent Graduate} stakeholder group as all students who graduated from the degree program in Fall 2021 or Spring 2022.  We collected data from them in the form of an online questionnaire, as documented in Appendix \ref{sec:graduating-student-questionnaire}.

We identified the {\em Faculty} stakeholder group as all tenure-track faculty in the computer science department as of Academic Year 2021-2022. We collected data from them in the form of an online questionnaire, as documented in Appendix \ref{sec:faculty-questionnaire}.

We identified the {\em Advisor} stakeholder group as all of the undergraduate advising staff in the computer science department as of Academic Year 2021-2022. We collected data from them in the form of an online questionnaire, as documented in Appendix \ref{sec:advisor-questionnaire}.

\subsection{Analysis}

\subsubsection{Data-Analytic Strategies}

For the Recent Graduate, Faculty, and Advisor stakeholder groups, the number of responses were too small (23, 10, and 2 respectively) to apply any formal statistical analysis.  Instead, we will present descriptive statistics and summaries of any short answer responses.

For the Early Stage Student stakeholder groups, we collected 187 essays, ranging in length from a short paragraph to several paragraphs. This was enough data to support a more formal analysis strategy of qualitative data coding \cite{braun_successful_2013}. In our case, we first generated a set of categories based upon our top-level research questions and preliminary review of a subset of essays, then randomly assigned each essay to two researchers to code independently. Our set of coding categories is listed in Appendix \ref{sec:coding-categories}. After each essay was coded by two researchers, we calculated Inter-Rater Reliability (IRR) for each encoding as the percentage of ratings that were in absolute agreement and found that we exceeded the "minimum standards" for reliability \cite{graham_measuring_2012} with an IRR of 79\%. We then performed a third review of codings in order to obtain a "tie-breaker" value for the percentage of essays including each coding category.

\section{Findings}

\subsection{Early Stage Student Stakeholders}

\subsubsection{Descriptive statistics}

As noted above, we collected data from Early Stage Student Stakeholders in the form of essays that they wrote soon after completing the New Student Tutorial.  Figure \ref{fig:essay-stats} provides an overview of this data. We collected data from a first semester class and a third semester class in both Fall and Spring semesters. We were only able to collect data from a second semester class in Spring semester. This resulted in a pool of 475 early stage students, of whom 39\% engaged in the New Student Tutorial and submitted an essay reflecting on their experience.

It is important when analyzing these findings to note that the essay prompt was non-specific: "a short 2 - 3 paragraph reflection on your experience with RadGrad."  This means that frequently occurring coding categories in essays provide significant evidence to support the assertion of the category, but it does not imply that infrequently occurring coding categories provide evidence {\em against} the assertion. For example, the essay writer might agree strongly with an assertion, but might have simply not thought to write about it or run out of time.

Figure \ref{fig:essay-stats} also provides descriptive statistics regarding the length of the essays in terms of the minimum, maximum, average, and median number of words. Interestingly, the length of the essays trends upwards with the semester: first semester students wrote the shortest essays, and third semester students wrote the longest essays.  A rule of thumb is that 500 words is equivalent to a page of text, and so the table reveals that first semester students wrote about a third of a page of text, while third semester students generally wrote about a page of text. There was a grand total of 48,993 words, or approximately 98 pages of essay writing to be coded.

\begin{figure}[th]
\centering
\small
\begin{tabular}{ l c c c | c c c c}
\hline
 Class & Total Students & Responses & Percentage & Min & Max & Ave & Median \\
\hline
 1XX, Fall 21 & 204 & 55 & 27\% & 22 & 433 & 156 & 146\\
 1XX, Spring 22 & 42 & 41 & 97\% & 61 & 370 & 185 & 165\\
 2XX, Spring 22 & 136 & 42 & 31\% & 98 & 499 & 262 & 256\\
 3XX, Fall 21 & 56 & 25 & 45\% & 188 & 828 & 460 & 432\\
 3XX, Spring 22 & 37 & 24 & 65\% & 249 & 877 & 467 & 433\\
\hline
Total & 475 & 187 & 39\% & 22 & 877 & 266 & 224\\
\hline
\end{tabular}
\caption{\em Essay Statistics. Min, Max, Average, and Median refer to the number of words in the essay.}
\normalsize
\label{fig:essay-stats}
\end{figure}

Figure \ref{fig:coding-stats} provides aggregate, descriptive statistics on the results of coding the 187 essays. The results are ordered by the percentage of essays containing comments that support the assertion of the coding category. To provide a deeper perspective on these numbers, the following sections present some quotes from the essays regarding the various categories.

\begin{figure}[th]
\centering
\small
\begin{tabular}{ p{4in} c c }
\hline
 Category &  Total & Percentage   \\
\hline
RadGrad provides useful information.                  & 143 & 76\% \\
The opportunity explorer is useful.                   & 140 & 75\% \\
The course explorer is useful.                        & 118 & 63\% \\
RadGrad broadened my perspective on computer science. & 117 & 63\% \\
The career goals explorer is useful.                  & 95 & 51\% \\
The degree planner is useful.                         & 85 & 45\% \\
The interests explorer is useful.                     & 67 & 36\% \\
RadGrad helps create communities of practice.         & 56 & 30\% \\
I will use RadGrad in future.                         & 41 & 22\% \\
RadGrad is well-designed and easy to use.             & 36 & 19\% \\
The game mechanics (ICE points, Levels) are useful.   & 31 & 17\% \\
Course and/or Opportunity Reviews are useful.         & 27 & 14\% \\
Internships are important in RadGrad.                 & 23 & 12\% \\
There are interactions between STAR and RadGrad.      & 17 & 9\% \\
RadGrad provides useful information
for women and/or underrepresented groups.             & 6 & 3\% \\
\hline
\end{tabular}
\caption{\em Coding Statistics, ordered by percentage. For each category, the number (and percentage) of essays that contained text agreeing with that assertion.}
\normalsize
\label{fig:coding-stats}
\end{figure}

\subsubsection{RadGrad provides useful information.} This is a kind of "aggregate" category, in that if an essay writer indicated that a specific RadGrad feature (Explorer, Degree Planner, Reviews, etc) was useful, then we infer that they also find RadGrad in general to be useful. As a result, we find that approximately three out of four essays included text that supported the assertion that RadGrad provides useful information. Here is an example excerpt that touches on many of the features:

\begin{itemize}[leftmargin=*]
\item "After spending about an hour going through RadGrad, I learned a lot more about courses, opportunities, and possible career interests. These are all things I hear older people talking about, but I never spend a lot of time thinking about them myself. I just let STAR tell me what classes to take, say I’ll think about opportunities later, and tell people I don’t know what I want to do in the future. RadGrad forced me to actually think about the classes I’m going to be taking. RadGrad also showed me opportunities available to me and made me choose some. RadGrad also made me choose computer science related jobs and see what common interests they share. Overall, in the hour or two I spent working on my RadGrad profile, I learned a lot about future classes, opportunities I can take advantage of, and several possible career choices."
\end{itemize}

\subsubsection{RadGrad Explorers are useful.} RadGrad includes four "explorers", all of which were mentioned frequently in essays. The Opportunity (i.e. extra-curricular activity) Explorer was mentioned in 75\% of the essays, the Course Explorer in 63\%, the Career Goals Explorer in 51\%, and the Interests Explorer in 36\%. Some excerpts supporting this assertion are:

\begin{itemize}[leftmargin=*]
\item "I thought that it was cool that they suggested some careers and extra-curriculars based on your interests and career goals.",
\item "Sometimes, college students have trouble finding extracurricular activities, and RadGrad organized, described, and allowed you to schedule these opportunities, which was very helpful.",
\item "At times I have felt rudderless, not knowing the correct classes to take for my degree plan. However, RadGrad reveals many different options with in-depth descriptions, allowing me to make an informed decision to further my degree path."
\end{itemize}

\subsubsection{RadGrad broadened my perspective on computer science.} After Opportunity and Course Explorer usefulness, the ability of RadGrad to "broaden one's perspective" was the most widely cited capability of the system, appearing in 63\% of the essays. Some excerpts supporting this assertion are:

\begin{itemize}[leftmargin=*]
\item "When I used to think about computer science, I would have a mental image of thousands of lines of code for some tech company to use in their newest product, for example a router, but I never realized how computer science could be used in so many non-technical fields. As someone who has always been extremely interested in social issues, I was really interested to hear about some of the applications and opportunities people are working on.",
\item "Previously, I thought that Computer science was just about writing code. Now that I saw RadGrad, I see that there are hundreds of interest categories available to focus on.",
\item "I have not known many careers in the computer field, the only three jobs I know when I was a freshman in college is software engineer, programmer, and web developer. In RadGrad, I learn a new career named ux designer, which is design web application based on users’ needs.",
\item "But what really impressed me was the career goals and courses. The career goals opened my eyes to the multitude of possibilities that ICS is capable of. Many of these career paths were unknown to me till that moment. I actually found a decent amount of careers that I wouldn’t mind exploring better."
\end{itemize}

There were also many comments associated with this category that implied a positive impact of RadGrad upon engagement and retention. For example,
\begin{itemize}[leftmargin=*]
\item "My first time I had to come to class I felt so lost and unsure of where I was going. I also do not know any of the clubs that are offered or the various opportunities that are available to us. Radgrad has helped me to get a better understanding of the different opportunities.",
\item "At that time, I was still on the edge of whether or not I wanted to commit to pursuing a degree in computer science. Therefore, seeing this website at that time helped me to see what kind of interests I had matched up with the courses or opportunities out there. At that time, it also helped take some of the uneasiness off my shoulders because I thought that going into computer science means there are only careers related to coding. I gained new insight into how much of a variety computer science can offer.",
\item "RadGrad is an entry way into a world of computer science I had no idea was within my reach. There were many new interests and career interests to be explored. I had no idea the number of events that go on throughout the year.",
\item "One class that I was scared of was ICS 311 which I was going to take, but a review urged me to want to take the class.",
\item "Lastly, seeing all of the opportunities available reminded me that I don't need to learn these things on my own--that there are tons of people at this University who are learning as well as can help me learn. Seeing all the opportunities on RadGrad has definitely made me feel a lot better and motivated to learn Computer Science.",
\item "Looking at RadGrad, knowing I'm not the best person at Computer Science, makes me feel less anxious about my possible future in the job field.",
\item "Going through and exploring the RadGrad page has made me become more interested in computer science and what the cs major actually does."
\end{itemize}

\subsubsection{RadGrad helps create communities of practice.} Approximately 30\% of the essays contained text that supported this assertion. As noted above, communities of practice can be a key driver for engagement and retention of women and under-represented groups. Some excerpts supporting this assertion are:

\begin{itemize}[leftmargin=*]
\item "Personally, I found the idea of discovering others with similar interests the most useful. I feel the most trouble I have is with networking and it was really cool to see some people I've seen in my classes with similar interests.",
\item "I always wanted to join a certain community and develop interaction skills with like-minded students. RadGrad provides me a various of communities that I can join and it makes a lot easier to see what community is the best fit for me.",
\item "Knowing that other people have similar interests and career focus as you creates a tightly knitted community."
\end{itemize}

\subsubsection{I will use RadGrad in future.}  Only 22\% of the essays were explicit that the author intended to use RadGrad in the future, even though 76\% of the essays were explicit that RadGrad was useful!  This may indicate the need that some additional incentives might be needed if RadGrad is to be used by significant numbers of students more than once. Some excerpts supporting this assertion are:

\begin{itemize}[leftmargin=*]
\item "I am 100\% going to use this resource sometime in the future",
\item "Knowing how helpful RadGrad can be now, I plan on using it for the rest of my attendance at [school]",
\item "I plan on using RadGrad regularly throughout my pursuit into the Software Engineer field.",
\item "I really hope to utilize RadGrad throughout my college career because it really is an amazing tool.".
\end{itemize}

\subsubsection{RadGrad is well-designed and easy to use.}  Approximately 19\% of the essays commented positively on the ease-of-use of RadGrad.
Some excerpts supporting this assertion are:

\begin{itemize}[leftmargin=*]
\item "I was surprised how easy RadGrad was to use. The design of it was very pleasing to the eye and easy to navigate. I also believe trying to rack up ICE points makes the experience a lot more enjoyable. Unlike other websites, I had an easy time understanding what each function did.",
\item "The design of the RadGrad site is visually appealing and the tutorial for completing a basic profile was easy to understand. The descriptions for the different topics of interest, career choices, ICS catalog courses, and opportunities were also well written.",
\item "I also like that the careers, interests, and courses all relate to each other and show a sort of ‘trajectory’, like showing what courses can apply to what interests and what interests apply to what careers.",
\item "With some more features I think this platform could be rolled out to other universities. While it might not be the intent, I think many university’s especially smaller ones could benefit.",
\item "I really enjoyed the snappy and stateful structure of the app, and the layout was pleasing and simple enough for easy navigation.",
\item "I'm really liking the styling you guys have done with RadGrad. Clean font and color palette. I also really like how convenient it is that it is connected to [the university's authentication system] so I don't need to go through the hassle and make an account before using RadGrad.
\item "The "What's New" section is really handy. This way if a new class or opportunity comes up then we'll be able to know."
\item "The RadGrad interface is very satisfying. It is intuitive, leaving little room for guessing or confusion."
\end{itemize}

However, there were a handful of negative comments regarding usability as well:
\begin{itemize}[leftmargin=*]
\item "It was hard to navigate around the site, the features were not very intuitive, and overall I just didn’t find myself getting much value from the application.",
\item "The RadGrad website seems a little clunky and cluttered. A lot of information is thrown to a new user and it can get confusing to follow."
\end{itemize}

\subsubsection{The game mechanics (ICE Points, Levels) are useful.} Just over 30\% of essays commented positively on RadGrad's use of game mechanics.  Some excerpts supporting this assertion are:
\begin{itemize}[leftmargin=*]
\item "One more thing I found very intriguing is the innovation, competency, and experience levels. The fact that RadGrad is pushing students to push themselves in all aspects of computer science is very awesome to me."
\item "The most unique thing about RadGrad is the ICE point system. It stands for Innovation, Competency, and Experience. It keeps track of the amount of points acquired from completed courses and opportunities. This is a useful system that helps students know what should be focused as they progress forward."
\item "I can feel my own growth with the increase in number(competency) which prevents me from developing imposter syndrome."
\item "RadGrad also offers a cool feature where there is a 'level' system where it showcases a student's level, and there are six levels in total for RadGrad...The level system that RadGrad has implemented can motivate and inspire students to continue to do well in their courses and in their extra-curricular activities."
\end{itemize}

\subsubsection{There are interactions between STAR and RadGrad.} The university's course registration system is called "STAR", which includes a degree planning mechanism. The essays revealed a spectrum of opinions regarding the interaction between RadGrad's degree planning environment and STAR's.  For example:
\begin{itemize}[leftmargin=*]
\item "It has very similar functionality like STAR except that it is more focus and oriented for CS students. It has a lot of features I wish STAR had. So, its presence is needed and justified. But at the same time, I wish I didn’t need to log into a separate website to use its tools.... My only wish is that STAR would implement the same type of features that Radgrad has."
\item "A planner that is 50000000x better than STAR is super awesome."
\item "RadGrad is a convoluted version of STAR."
\end{itemize}

\subsubsection{RadGrad provides useful information for women and/or underrepresented groups.} Unfortunately, explicit support for this assertion occurred in only 3\% of the essays. This is perhaps not surprising, given that we did not direct students to comment on this issue and it might not be top-of-mind for most of them. That said, there were a few comments supporting this assertion:

\begin{itemize}[leftmargin=*]
\item "I was also happy to see the Society of Women Engineers, as this industry is quite male-dominated and organizations like these are making the IT industry much less intimidating for women and encouraging them to partake."
\item I did not realize the number of opportunities [this institution] offered for women and minorities. I am particularly interested in the Society of Women Engineers. This is my first year and I haven't had many opportunities to meet like-minded individuals with similar interests.
\item "A club I will be joining is the SWITCH-UHM club. The club supports women in Information, technology and computing at the university and I never knew it existed before. I think it is important to be surrounded by people that support you and encourage you and I think it even more important to be around people that may be in the same situation such as the same CS classes... I look forward to joining this club and to meeting and making friends that are also fellow computer science chicas."
\item "Being a Computer Science major and a female I am already stressing out with making sure I am learning my material and proving that I am able to hold my own. However, with Radgrad I am able to feel a weight lifted off my shoulders knowing I have the resources I need to potentially succeed after graduation."
\end{itemize}

\subsection{Recent Graduate Stakeholders}

\subsubsection{Descriptive statistics.} We collected data from Recent Graduate stakeholders in the form of a questionnaire (see Appendix \ref{sec:graduating-student-questionnaire} for the list of questions). We contacted all 107 of these graduating students, and 23 of these students filled out the questionnaire, for a response rate of 21\%.  Figure \ref{fig:graduating-student-questionnaire-responses} summarizes the responses.

\begin{figure}[th]
\centering
\small
\begin{tabular}{ p{4in} c c }
\hline
 Questionnaire Response &  Total & Percentage   \\
\hline
I recall using RadGrad.                          & 20 & 87\% \\
I learned new things from RadGrad.               & 11 & 55\% \\
I used RadGrad "just once."                      & 4 & 20\% \\
I used RadGrad "2 or 3 times."                   & 11 & 55\% \\
I used RadGrad "4 or more times."                & 4 & 20\% \\
My use of RadGrad was affected by COVID-19.         & 4 & 20\% \\
The Interests Explorer was helpful.              & 9 & 45\% \\
The Career Goals Explorer was helpful.           & 10 & 50\% \\
The Courses Explorer was helpful.                & 12 & 60\% \\
The Opportunities Explorer was helpful.          & 13 & 65\% \\
The Reviews were helpful.                        & 7 & 35\% \\
The Degree Planner was helpful.                  & 6 & 30\% \\
The Levels and ICE Points were helpful.          & 3 & 15\% \\
\hline
\end{tabular}
\caption{\em Graduating Student questionnaire response descriptive statistics.}
\normalsize
\label{fig:graduating-student-questionnaire-responses}
\end{figure}

Out of 23 respondents, 3 indicated in the first question that they did not recall using RadGrad during their degree experience, and so only 20 respondents were asked the remaining questions.

We asked how often the respondents recalled using RadGrad. 55\% recalled using RadGrad "2 or 3 times", with 20\% using it 4 or more times. We checked for the impact of COVID-19, but only 20\% of respondents believed that it had an impact on their RadGrad usage.

One question asked respondents to check which RadGrad features they found helpful.  The most frequently checked feature was the Opportunity Explorer (65\%), followed by Courses, Career Goals, and finally Interests (at 45\%).  The feature that respondents found least useful was game mechanics (i.e. Levels and ICE points), which was checked by only 15\% of respondents.

55\% of respondents indicated that they learned new things from RadGrad. In half of these responses, Opportunities were cited as the new information. Reviews and the Degree Planner were also cited as helping respondents to learn new things.

When asked what we could do to make RadGrad more useful to future students, respondents mentioned: access to internships, greater promotion by faculty members, email notifications, and better integration with STAR.

\subsection{Faculty Stakeholders}

\subsubsection{Descriptive statistics.} We collected data from Faculty stakeholders in the form of a questionnaire (see Appendix \ref{sec:faculty-questionnaire} for the list of questions). We contacted all 21 instructional faculty, and 10 participated, for a response rate of 47\%.  Figure \ref{fig:faculty-questionnaire-responses} summarizes the responses.

\begin{figure}[th]
\centering
\small
\begin{tabular}{ p{4in} c c }
\hline
 Questionnaire Response &  Total & Percentage   \\
\hline
I recall reviewing the RadGrad Faculty tutorial. & 8 & 80\% \\
I recall logging into RadGrad.                   & 6 & 60\% \\
Does RadGrad contain Opportunities for your research?  & 6 & 60\% \\
\hline
\end{tabular}
\caption{\em Faculty stakeholder questionnaire response descriptive statistics.}
\normalsize
\label{fig:faculty-questionnaire-responses}
\end{figure}

The percentages in Figure \ref{fig:faculty-questionnaire-responses} are misleading, because the faculty who responded to this survey are also those who expressed some interest in RadGrad to the project team. The 11 faculty who did not participate in the survey also did not participate in RadGrad, so (for example) the percentage of faculty who logged into RadGrad is not best estimated as 6 out of 10 (60\%), but rather 6 out of 21 (28\%).  The faculty survey confirms what we observed anecdotally: faculty buy-in to RadGrad as a means of improving the undergraduate experience is low.

That said, the questionnaire did yield a few interesting Faculty perspectives on RadGrad:

\begin{itemize}[leftmargin=*]
\item "RadGrad has a lot of benefits for students/faculty/advisors. To minimize the range of systems, I think it would work well as a component of (or integrated with) STAR. That would simplify the location to 1-stop shop for educational endeavors."
\item "Many opportunity announcements have a short lifespan: people are looking for students to fill a position quickly, and once it is filled it's done. So ordering opportunities by most recent seems best (like discord). "
\item "The most pressing issue is faculty buy in.
\item "Make the use of RadGrad part of the advising and also part of the administration of the undergraduate programs. It should be used by the undergraduate committee members."
\item "I think we should adopt radgrad in CS in a more general way. My only worry is the maintenance of it. We should discuss how to have maybe a student help position just to do the maintenance? Is that enough?"
\end{itemize}

\subsection{Advisor Stakeholders}

\subsubsection{Descriptive statistics.} We collected data from Advisor stakeholders in the form of a questionnaire (see Appendix \ref{sec:advisor-questionnaire} for the list of questions). We contacted both undergraduate advisors, and both participated, for a response rate of 100\%.  Figure \ref{fig:advisor-questionnaire-responses} summarizes the responses.

\begin{figure}[th]
\centering
\small
\begin{tabular}{ p{4in} c c }
\hline
 Questionnaire Response &  Total & Percentage   \\
\hline
I recall reviewing the RadGrad Advisor tutorial. & 2 & 100\% \\
I recall logging into RadGrad.                   & 2 & 100\% \\
I usually use RadGrad in my advising sessions.  & 1 & 50\% \\
I always use RadGrad in my advising sessions.  & 1 & 50\% \\
RadGrad has a positive impact on student advising.  & 2 & 100\% \\
\hline
\end{tabular}
\caption{\em Advisor stakeholder questionnaire response descriptive statistics.}
\normalsize
\label{fig:advisor-questionnaire-responses}
\end{figure}

The very small number of Advisors makes it difficult to interpret statistics, but it is clear that these two advisors think RadGrad is useful for their purposes.

The questionnaire also produced some interesting comments on RadGrad from the Advisor perspective:

\begin{itemize}[leftmargin=*]
\item "I really like RadGrad as a database for students to look at opportunities and careers. It has helped students figure out what they want to do with their major."
\item "I use RadGrad for career exploration and opportunities mainly. I use STAR for everything else."
\item "I show them that RadGrad can help them explore and choose career paths and interests in computer science. I also show them the opportunities section and the courses section. I will tell them it may be helpful to review the syllabus and reviews from other students for a course before taking it to prepare."
\item "RadGrad can be a great starting point for students to explore the CS field and connect themselves to different opportunities (i.e. professional, internships, etc)."
\item "I think it helps students to think about things such as careers and interests early. The courses and opportunities section are also really good resources."
\end{itemize}

\subsection{Summary}

Figure \ref{fig:eval-questions-evidence} summarizes our findings by revisiting Figure \ref{fig:eval-questions}, this time providing a checkmark if the data appears to support the low-level research question, a question mark if the data does not provide evidence one way or another, and an X if the data appears to disconfirm the question.

\begin{figure}[th]
\centering
\small
\begin{tabular}{ p{1in} p{1in} c c c }
\hline
\multicolumn{2}{l}{Does RadGrad have a positive impact on...} & Engagement? & Retention? & Diversity? \\
\hline\\[0.05cm]
Are the following features useful... & &  &  & \\
\multicolumn{2}{r}{Explorers?} & \Checkmark & \Checkmark & \\
\multicolumn{2}{r}{Degree Planner?} & \Checkmark & \Checkmark & \\
\multicolumn{2}{r}{Game Mechanics?} & \Checkmark &  & \\
\multicolumn{2}{r}{Reviews?} & \Checkmark & \Checkmark & \\[0.5cm]
RadGrad is... & &  &  & \\
\multicolumn{2}{r}{well designed?} & \Checkmark &  & \\
\multicolumn{2}{r}{useful for women and underrepresented groups?} &  & ? & ? \\
\multicolumn{2}{r}{helpful for creating communities of practice?} & \Checkmark & \Checkmark & \Checkmark \\
\multicolumn{2}{r}{helpful for broadening student perspectives on CS?} & \Checkmark & \Checkmark & \Checkmark \\
\multicolumn{2}{r}{useful for faculty?} & X & X & \\
\multicolumn{2}{r}{useful for advisors?} & \Checkmark & \Checkmark & \\
\hline
\end{tabular}
\caption{\em Summary of findings: a "\Checkmark" indicates that data supported the low-level question; a "?" indicates the data did not provide evidence one way or the other, and an "X" indicates the data disconfirmed the low-level question.}
\normalsize
\label{fig:eval-questions-evidence}
\end{figure}

In general, our findings provided evidence to support the usefulness of all of the basic RadGrad features: Explorers, Degree Planner, Game Mechanics, and Reviews.  We found evidence that RadGrad is well designed, that it can help create communities of practice, and that it helps broaden student perspectives on computer science.  Both Advisors found RadGrad to be useful.

However, our data did not provide enough evidence to conclude that RadGrad is useful for women and underrepresented groups. We found evidence for this in only three essays, which we do not believe is sufficient for a conclusion. Thus, we leave this question as an open question.

We found disconfirming evidence that RadGrad is useful for Faculty, at least in the Department we studied for the time period of the study. Very few faculty engaged with the system, and despite the presence of a tutorial, very few seemed to understand or care about the importance of faculty engagement with RadGrad.

Finally, we believe the data for our low-level research questions provides fairly strong evidence that RadGrad can have a positive impact on engagement.  It provides somewhat less strong evidence that RadGrad can have a positive impact on retention.  It provides some indirect evidence that RadGrad can have a positive impact on diversity, since prior research shows that diversity efforts are more successful when there exist communities of practice and an understanding of the broader impacts of the discipline. We did not, however, obtain direct evidence for the positive impact of RadGrad on diversity.

\section{Discussion}

\subsection{Central Contributions to the Discipline}

Improving engagement, retention, and diversity of students in STEM disciplines in general and computer science in particular is important if we are to successfully address many important social, economic, and environmental challenges. The RadGrad Project provides a novel way to approach this issue that might be summed up in one word as {\em broadening}.  RadGrad broadens the student perception of progress metrics beyond GPA to a system that rewards Innovation, Competency, and (real-world) Experiences.  RadGrad broadens student perceptions of "curriculum" by effectively removing the "extra" from extra-curricular activities.  Finally, RadGrad broadens the range of approaches to institutional curriculum change.

Our case study provides qualitative evidence that RadGrad can support increased engagement and retention of students in computer science, and provides limited evidence that RadGrad can support an increase in diversity.

The project also contributes a high quality, tailorable, open source software framework that can be used by a wide variety of institutions. Institutions can use our software directly by downloading and installing the software, and then populating it with content appropriate to their discipline and location.  Institutions can also use our software indirectly, as a kind of "executable design" to help with implementation of alternative systems or initiatives.

\subsection{Limitations}

A significant limitation of this study is that we cannot ascribe any causal relationship between the evidence for engagement and retention and actual engagement and retention. For example, although retention in the department has increased slightly (from 81\% in 2017 to 85\% in 2021), our study does not provide evidence to support a causal connection to this increase.

Similarly, while the percentage of women in our program has increased slightly (from 17\% in 2017 to 24\% in 2021), our study cannot provide evidence that RadGrad had a role.

Indeed, our belief is that there is no "silver bullet" regarding engagement, retention, and diversity in STEM disciplines. Initiatives like RadGrad have a positive role to play, but they are most likely to be effective as part of a larger commitment within the institution.

\subsection{Implications for the Future}

After five years of design and development, we believe the RadGrad software system has now matured to a point where it can be evaluated by other institutions. Indeed, this case study provides evidence that the RadGrad system is highly usable. One student went so far as to recommend that it be adopted by other academic institutions.

One significant outcome from trial adoption in other departments is to see if our negative finding involving faculty stakeholders is true more generally. Ultimately, experience with a variety of departments and institutions can potentially lead to a set of heuristics that can help predict whether RadGrad can be successfully integrated into a department.

Longer term, it would be interesting to engage with an additional stakeholder group: the local community. RadGrad is designed to connect students to local communities of practice, and it would be interesting to see if those communities detect an influence of RadGrad on their membership and functioning.

\section{Acknowledgements}

(Omitted for the purposes of anonymous review.)

\section{Appendices}

\appendix

\section{RadGrad Onboarding Student Essay Prompt}
\label{sec:onboarding-student-essay-prompt}

RadGrad was introduced to students in the introductory computer science sequence through an extra credit home work assignment. The assignment asked the student to go through the New Student Tutorial, submit a screenshot as verification that they actually used the system, and then write a 2-3 paragraph essay reflecting on their experience.

Here is a sample of the instructions:

\begin{quotation}
RadGrad is an NSF-sponsored research project designed to help students learn more about computer science and help decide if this is a path you want to pursue.

\medskip

\noindent To get an introduction to RadGrad in just 60 seconds, please watch this video:

https://www.youtube.com/watch?v=71WV4nYaIoI

\medskip

\noindent For this homework assignment, please go through the RadGrad New User Tutorial:

https://www.radgrad.org/docs/users/new-student/overview

\medskip

\noindent When you are done, please login to Laulima to submit the following:

(1) A screen shot of your RadGrad Home Page after you’ve completed the New User Tutorial. (You should have no high priority checklist items, and your degree plan should be complete.)

(2) a short 2 - 3 paragraph reflection on your experience with RadGrad.
\end{quotation}

\section{RadGrad Graduating Student Questionnaire}
\label{sec:graduating-student-questionnaire}

\begin{enumerate}[leftmargin=*]
\item Do you recall ever using the RadGrad system?
\item About how many times did you login to RadGrad on average over the past two years?
\item Do you feel your use of RadGrad was affected by COVID? If so, please explain.
\item What aspects of RadGrad did you find to be useful (check all that apply): Interests Explorer, Career Goals Explorer Courses Explorer, Opportunities Explorer, Degree Planner, Community Page, Levels and ICE Points, Reviews of Courses and Opportunities.
\item Is there anything you learned from RadGrad that you didn't know before using it? If so, could you briefly describe it? (If not, just write "nothing")
\item What could we add to RadGrad to make it more useful to future students?
\item Is there anything else you'd like to tell us about RadGrad?
\end{enumerate}

\section{RadGrad Advisor Questionnaire}
\label{sec:advisor-questionnaire}

\begin{enumerate}[leftmargin=*]
\item Do you recall ever reviewing the RadGrad New Advisor Tutorial? \newline (https://www.radgrad.org/docs/users/new-advisor/overview)
\item Do you have any comments about the RadGrad New Advisor Tutorial?
\item Do you recall ever logging into the RadGrad instance for the ICS Department? \newline (https://radgrad2.ics.hawaii.edu/)
\item Do you have any comments about your experience using the RadGrad instance for the ICS Department?
\item How often do you use RadGrad in your advising sessions?
\item If you do use RadGrad in your advising sessions, please describe what you do.
\item If you don't use RadGrad in your advising sessions, why not?
\item Do you feel RadGrad can have a positive impact on student advising in the ICS Department?
\item If so, please describe the positive impact you believe it can have.
\item How could RadGrad (and/or Department procedures) be changed so that it makes a more positive impact on the undergraduate student experience?
\item Do you have any other comments regarding RadGrad?
\end{enumerate}

\section{RadGrad Faculty Questionnaire}
\label{sec:faculty-questionnaire}

\begin{enumerate}[leftmargin=*]
\item Do you recall ever reviewing the RadGrad New Faculty Tutorial? \newline (https://www.radgrad.org/docs/users/new-faculty/overview)
\item Do you have any comments about the RadGrad New Faculty Tutorial?
\item Do you recall ever logging into the RadGrad instance for the ICS Department? \newline (https://radgrad2.ics.hawaii.edu/)
\item Do you have any comments about your experience using the RadGrad instance for the ICS Department?
\item Does the ICS instance of RadGrad contain any Opportunities associated with your research?
\item If there are no Opportunities associated with your research, why not?
\item How could RadGrad (and/or Department procedures) be changed so that it makes a more positive impact on the undergraduate student experience?
\item Do you have any other comments regarding RadGrad?
\end{enumerate}

\section{Early Stage Student Essay Coding Categories}
\label{sec:coding-categories}

A preliminary review of the essays revealed the following themes of relevance to our research questions, which became the coding categories used in this study.  For each category, we provide an example excerpt from an essay that would cause that essay to be coded with a "1" for that category.

\noindent {\em RadGrad provides useful information.}
\begin{itemize}
\item "I was surprised by the sheer amount of information that I was able to find."
\end{itemize}

\noindent {\em The course explorer is useful.}
\begin{itemize}
\item "Being able to see all the possible ICS courses I could take was also particularly helpful."
\end{itemize}

\noindent {\em The opportunity explorer is useful.}
\begin{itemize}
\item "The opportunities tasks were also quite interesting."
\end{itemize}

\noindent {\em The career goals explorer is useful.}
\begin{itemize}
\item "I found it extremely useful to see which career goals fit my selected interests"
\end{itemize}

\noindent {\em The interests explorer is useful.}
\begin{itemize}
\item "My favorite part of the whole RadGrad experience was looking into what my interests are because I discovered a bunch of different skills/interests that I did not even know existed!"
\end{itemize}

\noindent {\em The degree planner is useful.}
\begin{itemize}
\item "I also like the degree planner because I got to read the descriptions of the different ICS classes that I can take. It also organizes the classes and opportunities by semester which I think is very helpful because it helps me visualize the classes I should be taking and the time I have."
\end{itemize}

\noindent {\em The game mechanics (ICE points, Levels) are useful.}
\begin{itemize}
\item "I also believe trying to rack up ICE points makes the experience a lot more enjoyable. "
\item "The ability to "level up" is personally enticing, as well."
\end{itemize}

\noindent {\em Course and/or Opportunity Reviews are useful.}
\begin{itemize}
\item "Being able to see other people’s reviews and see their experiences has also proven to be very useful in helping me decide which courses and opportunities I would like to try."
\end{itemize}

\noindent {\em RadGrad provides useful information for women and/or underrepresented groups.}
\begin{itemize}
\item I did not realize the number of opportunities that UH Manoa offered for women and minorities. "
\end{itemize}

\noindent {\em RadGrad is well-designed and easy to use.}
\begin{itemize}
\item "I was surprised how easy RadGrad was to use. The design of it was very pleasing to the eye and easy to navigate."
\end{itemize}

\noindent {\em I will use RadGrad in future.}
\begin{itemize}
\item "I can absolutely see myself using throughout my undergraduate experience."
\end{itemize}

\noindent {\em There are interactions between STAR and RadGrad.}
\begin{itemize}
\item "a planner that is 50000000x better than STAR is super awesome."
\end{itemize}

\noindent {\em RadGrad helps create communities of practice.}
\begin{itemize}
\item "What I also like about RadGrad is that we can find like-minded people to work on creation of projects together."
\end{itemize}

\noindent {\em Internships are important in RadGrad.}
\begin{itemize}
\item "I wish I could see the internship opportunities because one thing I learned from my siblings is that interning is very important."
\end{itemize}

\noindent {\em RadGrad broadened my perspective on computer science.}
\begin{itemize}
\item "RadGrad presented many opportunities and courses that I didn't even know existed."
\end{itemize}

\bibliographystyle{ACM-Reference-Format}
\bibliography{radgrad}

\end{document}